\documentclass[lettersize,journal]{IEEEtran}
\usepackage{amsmath,amsfonts}

\usepackage{algorithm}
\usepackage{array}
\usepackage[caption=false,font=normalsize,labelfont=sf,textfont=sf]{subfig}
\usepackage{textcomp}
\usepackage{stfloats}
\usepackage{url}
\usepackage{verbatim}
\usepackage{graphicx}
\usepackage{cite}

\usepackage{algpseudocode}
\usepackage{booktabs}
\usepackage{multirow}
\usepackage{enumerate}
\usepackage{amssymb}
\usepackage{amsmath}
\usepackage{marvosym}
\usepackage{url}
\usepackage{verbatim}
\usepackage{pifont}

\usepackage{stfloats}

\usepackage{xcolor}

\usepackage{epstopdf}
\begin{document}

\title{Simplicity over Complexity: An ARN-Based Intrusion Detection Method for Industrial Control Network}

\author{Ziyi Liu, Dengpan Ye,~\IEEEmembership{Member,~IEEE,} Changsong Yang, Yong Ding, Yueling Liu, Long Tang and Chuanxi Chen
\thanks{This work was supported by the National Key R\&D Program of China (2023YFB3107303), the Guangxi Natural Science Foundation (2024GXNSFAA010453, 2024GXNSFDA010064), the National Natural Science Foundation of China NSFC (No. 62472325 and 62072343), the Fundamental Research Funds for the Central Universities (No. 2042023kf0228),). (Corresponding author: Dengpan Ye)}
\thanks{Ziyi Liu, Dengpan Ye, Long Tang and Chuanxi chen are with the Key Laboratory of Aerospace Information Security and Trusted Computing, Ministry of Education, School of Cyber Science and Engineering, Wuhan University, Wuhan, 430072, China (e-mail: ziyi\_liu@whu.edu.cn; yedp@whu.edu.cn; l\_tang@whu.edu.cn; chencx@whu.edu.cn).}

\thanks{Changsong Yang, Yong Ding and Yueling Liu are with the Key Laboratory of Cryptography and Information Security, Guilin University of Electronic Technology, Guilin 541004, China, and also with the Guangxi Enginecring Rescarch Center of Industrial Internet Security and Blockchain, Guilin University of Electronic Technology, Guilin 541004, China (e-mail: csyang@guet.edu.cn; stone\_dingy@guet.edu.cn; ylliu@guet.edu.cn).}

}

\markboth{Journal of \LaTeX\ Class Files,~Vol.~14, No.~8, August~2021}%
{Shell \MakeLowercase{\textit{et al.}}: A Sample Article Using IEEEtran.cls for IEEE Journals}


\maketitle

\begin{abstract}
  Industrial control network (ICN) is characterized by real-time responsiveness and reliability, which plays a key role in increasing production speed, rational and efficient processing, and managing the production process. Despite tremendous advantages, ICN inevitably struggles with some challenges, such as malicious user intrusion and hacker attack. To detect malicious intrusions in ICN, intrusion detection systems have been deployed. However, in ICN, network traffic data is equipped with characteristics of large scale, irregularity, multiple features, temporal correlation and high dimensionality, which greatly affect the efficiency and performance. To properly solve the above problems, we design a new intrusion detection method for ICN. Specifically, we first design a novel neural network model called associative recurrent network (ARN), which can properly handle the relationship between past moment hidden state and current moment information. Then, we adopt ARN to design a new intrusion detection method that can efficiently and accurately detect malicious intrusions in ICN. Subsequently, we demonstrate the high efficiency of our proposed method through theoretical computational complexity analysis. Finally, we develop a prototype implementation to evaluate the accuracy. The experimental results prove that our proposed method has sate-of-the-art performance on both the ICN dataset SWaT and the conventional network traffic dataset UNSW-NB15. The accuracies on the SWaT dataset and the UNSW-NB15 dataset reach 95.48\% and 97.61\%, respectively.
\end{abstract}

\begin{IEEEkeywords}
Network Security, Industrial Control Network, Intrusion Detection, Associative Recurrent Network.
\end{IEEEkeywords}

\section{Introduction}
\IEEEPARstart{W}{ith} the rapid development and integration of industrial informatization and automation, industrial control system (ICS) has been widely applied, which connects plenty of industrial infrastructures and equipment through industrial control network (ICN)~\cite{Galloway2012Introduction}. Generally, ICN is characterized by real-time responsiveness and reliability, which plays a key role in production acceleration and production process management. Recently, due to the widespread popularity of 5G wireless network~\cite{Cao2020Reliable,Koursioumpas2021AIdriven}, cloud computing and artificial intelligence, ICN has been widely used in natural gas pipeline systems, power transmission systems, industrial power generation systems, etc. As a result, both governments and enterprises are paying increasing attention to ICN.

Despite tremendous advantages, ICN inevitably struggles with some security and technical challenges, such as network security protection. In ICN, there are a large number of distributed industrial infrastructures and equipment~\cite{Ding2019Asurvey}. However, the security of ICN physical isolation is broken due to the fusion and development of information technology~\cite{Lv2020Trust}, thus bringing a series of new security problems. Meanwhile, ICN might be frequently invaded by malicious users and hackers, which not only causes damage to ICN but also poses a threat to the entire ICS. Therefore, the security protection of ICN is particularly important. However, most of the traditional network security defense methods, such as firewall and vulnerability scan, can only achieve passive defense.

To achieve active defense in ICN, intrusion detection~\cite{} technology has been widely applied. Intrusion detection is a typical active defense technology that aims to detect network intrusions or intrusion attempts by analyzing network traffic data, such as user behavior data, behavior log data and audit data. By deploying intrusion detection systems, network security managers can monitor the entire ICN without affecting the network performance. Hence, network security managers can easily understand the network security situation and take related measures, thereby providing ICN devices and systems with real-time security protection. Therefore, intrusion detection attracts lots of attention, and has produced a large number of theoretical and engineering results. Roughly, the existing intrusion detection methods can be divided into three categories: statistics-analysis-based intrusion detection, time-series-based intrusion detection and deep-learning-based intrusion detection.

Although plenty of intrusion detection~\cite{Du2023Afewshot,He2024Reinforcement} methods have been proposed, there are still some internal deficiencies when they are directly applied in ICN. Firstly, ICN traffic data is characterized by high dimensionality and large scale. As a result, statistics-analysis-based intrusion detection methods require plenty of mathematical calculations to process these data, resulting in expensive computational overhead and detection delay. Secondly, Time series-based intrusion detection methods usually focus on the temporal attributes of data, and rarely include other data attributes in the scope of analysis. However, except the temporal correlation, there are still many factors that might greatly affect the intrusion detection result. Therefore, time-series-based intrusion detection methods are degraded by low accuracy in ICN. Last but not least, note that ICN traffic data is also equipped with properties of diversity, unbalanced samples and temporal correlation. However, traditional deep-learning-based intrusion detection methods cannot perfectly handle the above properties, thus reducing the generalization and accuracy.

Recurrent neural network (RNN)~\cite{Wu2021Tensor,Jha2020Recurrent} can learn and characterize ICN traffic data, thus providing potential solution to address the above problems~\cite{Abdel2020DEEPIFS}. Therefore, the motivation of this paper is to design a new intrusion detection method for ICN. Specifically, we aim to improve the traditional gated recurrent unit (GRU)~\cite{ChoK2014} and design a novel neural network model, which can properly process the ICN traffic data. Subsequently, we aim to adopt the new designed neural network model to propose a novel intrusion detection method, which can realize precise and efficient intrusion detection in ICN.

\subsection{Contributions}
In this paper, we investigate a fundamental but challenging issue, i.e., how to improve the accuracy and efficiency of intrusion detection in ICN. Specifically, we design a novel RNN model, namely, associative recurrent network (ARN). Then, we adopt ARN to propose a new intrusion detection method for ICN. Therefore, the main contributions of this paper can be summarized as the following three folds.


$\bullet$ We first design a new recurrent neural network called ARN for intrusion detection in industrial control networks. Compared with GRU models, ARN eliminates the conflict risk between gated units and can better characterize the long-term and short-term dependencies of data.


$\bullet$ We design a single attention (S-ATT) mechanism to retain the past hidden state information in ARN. S-ATT uses the attention mechanism to directly learn the correlation between the past hidden state information and the current input information, allowing ARN to abandon the traditional gated unit structure and more reasonably preserve past moment hidden state information.

$\bullet$ We compare the theoretical computational complexity of our proposed ARN-based intrusion detection method and the GRU-based intrusion detection method. The comparison results demonstrate that they have the same theoretical computational complexity. Moreover, we design a prototype system and simulate our proposed method to evaluate the performance. Experimental results show that the accuracies on the SWaT dataset and the UNSW-NB15 dataset reach 95.48\% and 97.61\%, respectively.

\subsection{Related work}
Generally, the existing intrusion detection methods can be roughly summarized into three categories: statistics-analysis-based intrusion detection method~\cite{Aitchison1982}, time-series-based intrusion detection method~\cite{Zou2019} and deep-learning-based intrusion detection method~\cite{Dong2021As}.

\textbf{Statistics-analysis-based intrusion detection method.} Wu et al.~\cite{Wu2018} proposed an information entropy intrusion detection system (IDS) based on improved sliding window strategy. By improving the optimizing conditions of the decision and sliding window strategy, their method could enhance the IDS accuracy. To detect invisible attacks in industrial control systems, Hu et al.~\cite{Hu2020Detecting} represented the ordered content in the residual by permutation entropy. According to certain rules presented by sensor measurement values Zhou et al.~\cite{Zhou2021Permutation} proposed an industrial information system attack detection scheme based on permutation entropy. Liao et al.~\cite{Liao2022} designed a gray-entropy-based IDS, which could analyze and cluster the traffic gray level. Meanwhile, their designed system had higher intrusion detection accuracy for both known and unknown traffic attacks. However, the above entropy-based intrusion detection methods rely on distribution assumptions and involve a lot of complex mathematical calculations. When there is a lot of noise in industrial network traffic and the traffic scale is large, they are difficult to detect attack traffic.


Liang et al.~\cite{Liang2019An} divided the data weighted distance and safety factor of each node in the network according to the priority threshold of data attribute characteristics, and designed an ICN intrusion detection method based on multi-feature data clustering optimization model. Mittal et al.~\cite{Mittal2022A} implemented a new clustering method using a new variant of the gravitational search algorithm and used it for detection in the Industrial Internet of Things (IoT). However, clustering algorithms usually require a lot of computational overhead and memory usage overhead when facing large-scale data. Therefore, when facing large-scale industrial network data, clustering-algorithm-based intrusion detection methods are difficult to meet the defense needs of the industry.

In summary, we can see that statistics-analysis-based intrusion detection methods require many mathematical calculations and analyses, resulting in a heavy computational burden. Meanwhile, while processing high-dimensional and irregularity data, the accuracy of the above methods was insufficient.


\textbf{Time-series-based intrusion detection method.} Bozdal et al.~\cite{Bozdal2021} presented a wavelet-based ingress method for controller area network (CAN). By analyzing changes in network traffic data, their proposed method could precisely determine attack patterns. To describe the intrusion attack, Miao et al.~\cite{Miao2020} investigated the attack signals of ICS intrusions. By conducting frequency domain analysis on network traffic, they obtained the influence of attack signal frequency domain on the estimation performance of linear attack signal estimator. Fouladi et al.~\cite{Fouladi2016} adopted frequency-based discrete wavelet transform and discrete Fourier transform to design a naive Bayesian classifier, which could exactly distinguish attack traffic data from normal traffic data. However, the above wavelet-based detection methods are limited by noise interference and the selection of characteristic signals, resulting in detection deviation.

Ye et al.~\cite{Ye2013} adopted Hurst parameter estimation to design a fractional Fourier transform (FRFT) intrusion detection method. At the same time, they developed a prototype implementation to test the performance. Experimental results showed that their proposed method was not affected by nonstationary time series in Hurst estimation. Huang et al.~\cite{Huang2020False} simultaneously extracted time domain and frequency domain features from the sensor measurement data required for closed-loop control of industrial control systems, and used the features to establish a hidden Markov model to implement intrusion detection. However, the Fourier-transform-based detection methods are susceptible to boundary effects and the hidden Markov transform is limited by the dependence on the previous moment. Therefore, it is difficult to achieve accurate industrial network intrusion detection.

In addition, time-series-based intrusion detection method mainly use the time attribute of data as an analysis factor, and rarely study the influence of other external factors, which seriously limits their scope of application in intrusion detection.

\textbf{Deep-learning-based intrusion detection method.} Li et al.~\cite{Li2020Robust} used a multi-convolutional neural network (multi-CNN) fusion method to perform intrusion detection on industrial IoT traffic. Abdelaty et al.~\cite{Abdelaty2021DAICS} constructed a deep learning anomaly detection in ICS (DAICS) using a 2-branch deep neural network (DNN). Their method learned the changes in the behavior of industrial control systems through a few data samples and gradient updates, thus effectively improving the accuracy and robustness of intrusion detection. Yang et al.~\cite{Yang2019} combined levenberg-marquard (LM) and back propagation neural network (BPNN) to present a new IDS. In their proposed method, LM was utilized to optimize the weight of BPNN, thus improving accuracy. Sood et al.~\cite{KSood2023} proposed a two-stage IDS, which could achieve data dimensionality reduction, and was compatible with the ETSI-NFV standard 5G architecture. Experimental results showed that their method has higher accuracy in traffic detection. However, the above deep learning methods fail to utilize the temporal sequence between network traffic, resulting in poor results in intrusion detection.

Jadidi et al.~\cite{Jadidi2023Multi} proposed a causal anomaly detection method for industrial control systems, which consists of two stages. First, attack detection is performed through a classifier with LSTM as the core, and then the causal relationship between ICS logs is analyzed to diagnose the future impact of the attacker. Xu et al.~\cite{Xu2018} utilized GRU, multilayer perceptron (MLP) and softmax layer to design a new IDS.  Zare et al.~\cite{Zare2024A} designed an autoencoder with long short-term memory units, teacher forcing technique, and attention mechanism for IDS in ICS. However, most of the RNN-based intrusion detection methods have mutual influence between gating units, which makes it difficult to effectively manage the potential connections between long-term and short-term dependencies of data, limiting the accuracy of intrusion detection.

In summary, although the deep-learning-based intrusion detection methods can effectively characterize large-scale network traffic data, there is still an urgent need for a new method that is not affected by the conflict of gated units and can effectively use the temporal relationship between networks for representation learning to improve the performance of intrusion detection. To this end, we will conduct in-depth research on this issue and propose a new deep-learning-based intrusion detection method. We construct Table~\ref{tab: relatework} to summarize the comparison between intrusion detection solutions based on deep learning.

\begin{table}[]

\caption{Comparison of related research on intrusion detection methods based on deep learning.}

\resizebox{\columnwidth}{!}{

\begin{tabular}{cccc}

\hline
Method                                         & Whether Temporal Learning   &   No Gate Conflict \\ \hline
Li et al.~\cite{Li2020Robust}                    & \ding{55}                                                               & $\checkmark$                                                                \\
Abdelaty et al.~\cite{Abdelaty2021DAICS}                  & \ding{55}                                                               & $\checkmark$                                                                \\
Yang et al.~\cite{Yang2019}                     & \ding{55}                                                               & $\checkmark$                                                                \\
KSood et al.~\cite{KSood2023}                     & \ding{55}                                                               & $\checkmark$                                                                \\
Jadidi et al.~\cite{Jadidi2023Multi}                     & $\checkmark$                                                              & \ding{55}                                                               \\
Xu et al.~\cite{Xu2018}                   & $\checkmark$                                                               & \ding{55}                                                               \\
Zare et al.~\cite{Zare2024A}                    & $\checkmark$                                                             & \ding{55}                                                               \\
\textbf{Ours}                                & \textbf{$\checkmark$}                                                     & \textbf{$\checkmark$ }                                                     \\ \hline
\end{tabular}}

\label{tab: relatework}
\end{table}

\subsection{Organization}
The rest of this paper is organized as follows. In Section~\ref{sec:preliminaries}, we provide a detailed introduction to GRU and single attention (S-ATT). Subsequently, in Section~\ref{sec:arn} and Section~\ref{sec:method}, we present the structure of ARN and design an ARN-based intrusion detection method, respectively. Then, in Section~\ref{sec:analysis}, we provide a theoretical computational complexity comparison of the ARN-based intrusion detection method and GRU-based intrusion detection method. Next, we develop a prototype implementation for our proposed method and present the performance evaluation in Section~\ref{sec:implement}. Finally, we provide a brief conclusion in Section~\ref{sec:con}.

\section{Preliminaries}
\label{sec:preliminaries}

In this following section, we introduce some preliminaries that will be utilized to establish our new intrusion detection method, including GRU and S-ATT.

\subsection{Gated recurrent unit}
\label{sec:gru}

\begin{figure}[h]
  \centering
  \includegraphics[width=3.5in]{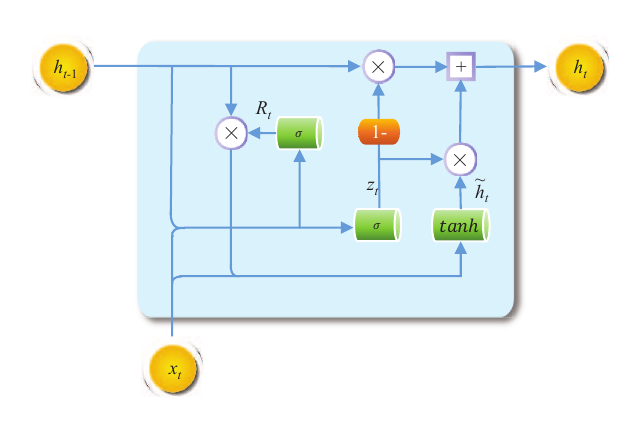}
\caption{The structure of GRU}
  \label{GRU}
\end{figure}

Gated recurrent unit (GRU) is an improved model of RNN that can effectively transmit long-term dependency information. In contrast to long short-term memory (LSTM), which is equipped with three gate structures, GRU only contains two gates: the reset gate and the update gate, as illustrated in Figure~\ref{GRU}. As a result, GRU requires fewer parameters and can reduce the risk of overfitting during training. In GRU, the reset gate is utilized to decide the proportion of hidden state information that flows into the current candidate set at the previous moment. Meanwhile, the update gate has two functions. On the one hand, it is utilized to control the proportion of the content that should be discarded at the previous moment. On the other hand, it can also determine the proportion of the content that needs to be retained in the current candidate set.

As shown in Figure~\ref{GRU}, the reset gate is denoted by $R_t$, the update gate is denoted by $Z_t$, the weights for update and reset gates are $W_Z$ and $W_R$, the candidate set is denoted by $\tilde{h_t}$, the current hidden state is denoted by $h_t$ and the past hidden state is denoted by $h_{t-1}$. The related formulas for calculating GRU are as follows.

Firstly, $R_t$ and $Z_t$ must be initialized according to Formula (1) and Formula (2) as follows:

\begin{equation}
  R_t=\sigma([h_{t-1},x_t]\cdot W_R)
\end{equation}

\begin{equation}
  Z_t=\sigma([h_{t-1},x_t]\cdot W_Z)
\end{equation}

Secondly, $\tilde{h_t}$ of GRU is controlled by $R_t$. Meanwhile, $\tilde{h_t}$ can be described by Formula (3) as follows:

\begin{equation}
  \tilde{h_t} = tanh(W_{\tilde{h}}\cdot[x_t,R_t \times h_{t-1}])
\end{equation}

Finally, GRU updates $h_t$ according to Formula (4):

\begin{equation}
  h_t = Z_t\times \tilde{h_t}+(1-Z_t)\times h_{t-1}
\end{equation}

From the above formulas and Figure~\ref{GRU}, we find that the $Z_t$ and the $R_t$ conflict in the retention proportion of the hidden state information in the past. Specifically, the reset of the current hidden state in GRU is provided by the $R_t$ and $Z_t$. The role of the $R_t$ is to determine the proportion of the past hidden state retained in the current candidate set. However, the $Z_t$ can ignore part of the past hidden state that the $R_t$ has determined to retain. Meanwhile, the $Z_t$ arbitrarily retains the past hidden state in the process of resetting the current hidden state. Therefore, there is a certain conflict between the $Z_t$ and the $R_t$, making it difficult to effectively manage the retention relationship between the past moment hidden state and the current moment input.

\subsection{Single attention}
\label{sec:sa}

\begin{figure}[h]
  \centering
  \includegraphics[width=3.5in]{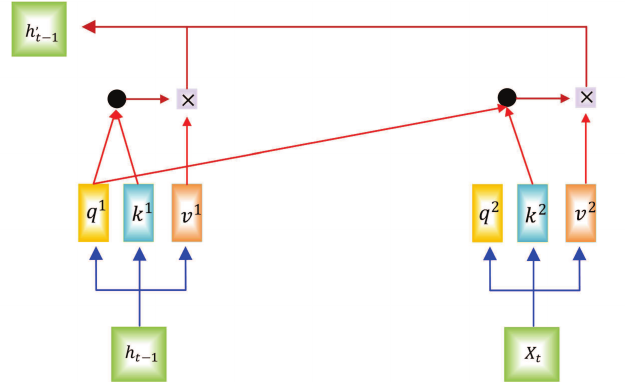}
  \caption{The structure of S-ATT}
  \label{ss}
\end{figure}

To effectively solve the problem of how to determine the retention relation between the hidden state of the past moment and the input of the current moment, we use the idea of Self-attention~\cite{Vaswani2017} to propose a one-way self-attention mechanism solution called single attention (S-ATT). S-ATT can effectively manage the retention relationship between the past hidden state (represented by $h_{t-1}$) and the current moment data (represented by $X_t$ represents the retained relationship between). The specific implementation process of the S-ATT method is shown in Figure~\ref{ss}.

As demonstrated in Figure~\ref{ss}, the $h_{t-1}^{\prime}$ represents the supplementary matrix, the query of $h_{t-1}$ and $X_t$ is represented by $q^1$ and $q^2$, the key of $h_{t-1}$ and $X_t$ is represented by $k^1$ and $k^2$, and the value of $h_{t-1}$ and $X_t$ is represented by $v^1$ and $v^2$. Then, we use $W^q$, $W^k$, and $W^v$ represent the weight matrices of $q$, $k$ and $v$, respectively. The symbol $a_{h}$ and $a_{x}$ represents the weight obtained by calculating the similarity between $q^1$ and each input $k$. Symbol $d$ represents the dimensions of $q$, $k$ and $v$, which are utilized to reduce the size of matrix $a_{h}$ and $a_{x}$ to ensure the stability of the gradient during back propagation. It is worth noting that the symbolic representation of this section still applies to Section~\ref{sec:arn} and Section~\ref{sec:method}.

First, we initialize $q^n$, $k^n$, $v^n$ (n=2) to facilitate subsequent calculations.
\begin{equation}
  q^1 = W^qh_{t-1},  q^2 = W^qX_t
\end{equation}

\begin{equation}
  k^1 = W^kh_{t-1},  k^2 = W^kX_t
\end{equation}

\begin{equation}
  v^1 = W^vh_{t-1},  v^2 = W^vX_t
\end{equation}

Then, we calculate the correlation between $h_{t-1}$ and $X_t$, and get the corresponding weight matrices $a_h$ and $a_x$.

\begin{equation}
  a_h = \frac{q^1 \cdot k^1}{\sqrt{d}},  a_x = \frac{q^1 \cdot k^2}{\sqrt{d}}
\end{equation}

\begin{equation}
  \widehat{a}_h, \widehat{a}_x = softmax(a_h, a_x)
\end{equation}

Finally, we calculate and fuse it into a new supplementary matrix $h_{t-1}^{\prime}$ based on the proportion of weights. In the following formula, $\widehat{a}_hv^1$ represents $h_{t-1,x}$, and $\widehat{a}_xv^2$ represents $X_{t,h}$.

\begin{equation}
  h_{t-1}^{\prime} = \widehat{a}_hv^1 + \widehat{a}_xv^2
\end{equation}

\section{Associative recurrent network}
\label{sec:arn}

\begin{figure}[t]
  \centering
  \includegraphics[width=3.5in]{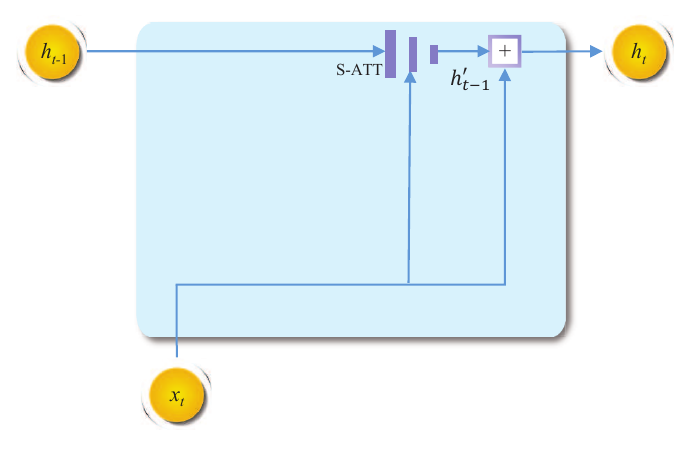}
  \caption{The structure of ARN}
  \label{ran}
\end{figure}

To address the problems of GRU that described in Section~\ref{sec:gru}, we propose a new RNN method called ARN. Generally, ARN utilizes S-ATT to unidirectionally learn the relationship between the past hidden state information and the current moment input information. As a result, ARN can construct an information complement matrix to complement the current moment input information and achieve the reset of the current hidden state information. The reset current moment hidden state includes important past information parts that are unknown in the information at the current moment and highlights the parts of the current moment information that are associated with the past hidden state. Therefore, ARN can better deal with the retention and relationship between past hidden states and current moment information, thereby avoiding gate conflicts existing in GRU and more effectively achieving intrusion detection of industrial control security traffic. The Figure~\ref{ran} shows the specific structure of ARN.

Firstly, we require to initialize $X_t$ and $h_{t-1}$ with $W^x$ and $W^{h_{t-1}}$ while making $X_t$ and $h_{t-1}$ have the same matrix dimension, which is convenient for subsequent correlation calculation. Then, $X_t$ and $h_{t-1}$ can be initialized as Formula (11) and Formula (12), respectively.

\begin{equation}
  X_t= W^xX_t
\end{equation}

\begin{equation}
  h_{t-1}= W^{h_{t-1}}h_{t-1}
\end{equation}

Secondly, we use stacks to splice $h_{t-1}$ and $X_t$ together to construct S-ATT input data, as shown in Formula (13).

\begin{equation}
  h_{stack}= stack(h_{t-1}, X_t)
\end{equation}

Subsequently, we utilize S-ATT to learn the relationship between $h_{t-1}$ and $X_t$, thus constructing a new complementary matrix $h_{t-1}^{\prime}$. The matrix $h_{t-1}^{\prime}$ contains part of the highlight important content in the current moment information, and the contents of hidden states in past moments that are unknown at the current moment, as shown in Formula (14).

\begin{equation}
  h_{t-1}^{\prime}= S-ATT(h_{stack})
\end{equation}

Finally, we supplement $h_{t-1}^{\prime}$ to $X_t$ to achieve the reset of $h_t$. The reset $h_t$ contains past time content that is unknow in the input data $X_t$ at the current time and highlight the important content in the current moment information, so as to better manage the retention relationship between past moment hidden state data and current moment data. Then, $h_t$ can be computed by using Formula (15).

\begin{equation}
  h_t=  h_{t-1}^{\prime} + X_t
\end{equation}

In summary, ARN can effectively learn the highlighted important parts of the current moment content in feature learning by constructing a supplementary matrix and fully consider the impact of the unknown past moment hidden state content on the current moment.

\section{ARN-based intrusion detection method}
\label{sec:method}

In the following section, we utilize ARN to build a new intrusion detection method for ICN.

\subsection{System model}
In this paper, we mainly study the creation of intrusion detection method for ICN. Note that with the rapid development of ICN, the application and popularity of ICN continuously increase. As a result, network traffic data is equipped with the characteristics of large scale, multiple features, temporal correlation, irregularity and high dimensionality, which pose some great challenges to achieve intrusion detection. Specifically, these characteristics directly affect the universality and accuracy of the intrusion detection method.

\begin{figure}[!h]
  \centering
  \includegraphics[width=3.7in]{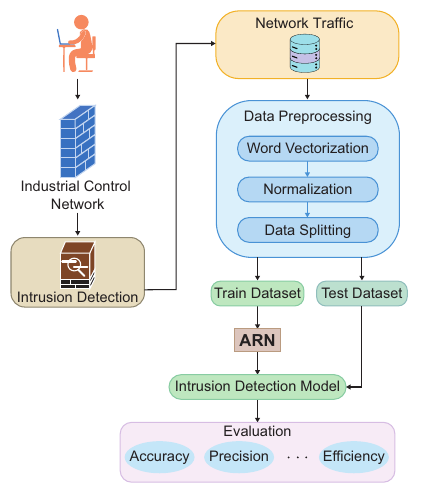}
  \caption{The system model}
  \label{method}
\end{figure}

To address the above problems, we propose a new intrusion detection method, as illustrated in Figure~\ref{method}. Specifically, a novel RNN model called ARN is proposed. Compared with other DNN models, ARN can not only learn and characterize network traffic data through time series but also better handle the retention relationship between past moment hidden state data and current moment input data. Subsequently, we utilize ARN to build a new intrusion detection method for ICN. By taking the advantage of ARN, our novel proposed method can greatly enhance the intrusion detection efficiency and accuracy.

\subsection{The proposed method}
In this section, we introduce the implementation of ARN-based intrusion detection method for ICN in detail.

To describe our proposed method more clearly, we firstly define some symbols. Specifically, the training dataset is denoted by $X-train$, the test dataset is denoted by $X-test$, the intrusion detection result is denoted by $S$, the data processing epoch is represented by $d$, the training epoch is denoted by $n$ , the real label is denoted by $Y_{real}$, the detection label is denoted by $Y_{pred}$, the comparison between the classification label and the true label is denoted by $Compare$, the performance evaluation of the trained model is denoted by $Evaluate$, and the evaluation result is denoted by $R$. Subsequently, we describe the process of our proposed intrusion detection method through pseudocode, as demonstrated in Algorithm 1.

\begin{algorithm}[htbp]
\caption{ARN-based intrusion detection method}
\hspace*{0.02in} {\bf Input:}
$X-train$, $Y-test$, $R-L$, $n$\\
\hspace*{0.02in} {\bf Output:}
result $S$
\begin{algorithmic}[1]
\State Step 1:Network traffic capture and preprocessing
\State Obtain industrial control security traffic data
\State Deliver traffic to an intrusion detection system (IDS)
\State Data Preprocessing
\For{$i=1 \to d $}
\State Word Vectorization
\State Normalization
\EndFor
\State Step 2: Data Splitting
\State Divide the processed data set into $X-train$ and $Y-test$.
\State Step 3: Model Training
\For{$i=1 \to n $}
\State $ARN \leftarrow ARN_i(X-train)$
\State $result \leftarrow Compare(Y_{real},Y_{pred})$
\EndFor
\State Step 4: Model test
\State $S \leftarrow ARN(X-test)$

\State $R \leftarrow Evaluate(S)$

\end{algorithmic}
\end{algorithm}

The implementation process of Algorithm 1 is as follows.

\begin{itemize}

  \item Step 1: Obtain the industrial control security traffic data, this part of the traffic data before accessing the ICS will be sent to the ARN intrusion detection method for detection, if found threats to the traffic data will be intercepted, otherwise the traffic data will be allowed to access;

  \item Step 2: Preprocess the data, including word vectorization of characters in the data, normalization of the data, etc. Then, the data is divided into training set $X-train$ and test set $Y-test$;

  \item Step 3: We utilize the training set to train the intrusion detection model. Among them, our proposed ARN method is the core of the intrusion detection model, which can effectively process ICN security traffic data. The specific implementation ideas of ARN are described as fellow:

      When the traffic data at the current moment is fed into the ARN, we first calculate the correlation between $h_{t-1}$ and $X_t$, and use the idea of Self-attention to query how much content in $h_{t-1}$ is known to $X_t$ (represented by $X_{t,h}$). This part is very important in time series data since this part of $X_t$ contains both the content input at the current moment and the connection with the past hidden state. Then, the remaining part of $h_{t-1}$ is the past information unknown to $X_t$ (represented by $h_{t-1,x}$). Subsequently, we combine $X_{t,h}$ and $h_{t-1,x}$ to form the supplementary matrix $h_{t-1}^{\prime}$. Finally, when the supplementary matrix is added to $X_t$, it can highlight the important content in $X_t$ and increase the supplementing of past information unknown to $X_t$. We compare the labels output by ARN with the true labels for model training.We compare the labels output by ARN with the true labels for model training, the trained model will be used for ICN intrusion detection;

  \item Step 4: The test set is fed into the trained intrusion detection model for evaluation. The ICS determines whether the traffic can access it based on the evaluation results.

\end{itemize}

\section{Computational complexity analysis}
\label{sec:analysis}

In this following section, we compare the theoretical computational complexity of our proposed ARN-based intrusion detection method with that of the GRU-based intrusion detection method, as demonstrated in Table~\ref{tab1}.

\begin{table}[t]
  \small\sf\centering
  \caption{Time complexity comparison.\label{T1}}
  \setlength{\tabcolsep}{1mm}{
  \begin{tabular}{ccc}
  \toprule
  & Computational overhead  &   Time complexity \\
  \midrule
  ARN& $T(2\times m\times n+12\times n^2 + 2\times n)$ & $O(n^2)$ 	\\
  GRU& $T(3\times m\times n + 6\times n^2+n)$ & $O(n^2)$ 	\\
  \bottomrule
  \end{tabular}}\\[10pt]
  \label{tab1}
\end{table}

For simplicity, we assume that the number of hidden units is $n$ and the dimensionality of the input data is $m$. Subsequently, the theoretical computational overhead for input and past hidden state initialization is $T(2\times m\times n)$, and the theoretical computational overhead of self-attention is $T(j\times (6n^2+n))$, where $j$ represents the number of matrices that need to learn the correlation. Since there are only 2 matrices that need to calculate the correlation in our proposed method, the theoretical computational overhead of self-attention is $T(12\times n^2 +2\times n)$. Moreover, the theoretical computational overhead for the current hidden state is $T(n)$, the total theoretical computational overhead of ARN is $T(2\times m\times n+12\times n^2 + 2\times n)$, and the theoretical time complexity is $O(n^2)$.

For GRU, the theoretical computational overhead of the two gates is $T(2\times m\times n+2\times n^2)$, the theoretical computational overhead of the candidate set is $T(m\times n+ 2\times n^2)$, and the theoretical computational overhead of the current hidden state is $T(2\times n^2 +n)$. Therefore, the total theoretical computational overhead is $T(3\times m\times n + 6\times n^2+n)$, and the theoretical time complexity is also $O(n^2)$.

\section{Scheme implementation}
\label{sec:implement}

In this following section, we first explicit the experimental environment, dataset and evaluation criteria in detail. Subsequently, we provide the experimental results and analysis.

\subsection{Experimental environment and dataset}
In this paper, all of the simulation experiments are carried out on a desktop equipped with a CUDA10.0 driver, cuDNN7.4.2 driver and Windows 10 operating system. Meanwhile, the desktop is equipped with 16 GB main memory, an NVIDIA GeForce GTX 2060S graphics card and an AMD Ryzen 5 3500X 6-core CPU that runs at 3.60 GHz. In addition, the experimental results of our proposed solution and comparative experiment are obtained by running in our experimental environment, the framework used in this experiment is based on pytorch1.5, the optimization function is Adam, the learning rate is 0.001, the network layer number of ARN is 1 and the number of hidden layer neurons is 100.

In our simulation experiments, the chosen datasets are SWaT and UNSW-NB15.

\textbf{SWaT} is a public safety dataset for industrial water treatment platforms~\cite{JGoh2016}. We chose the officially labeled ``SWaT Dec 2015" for the experiments. SWaT dataset consists of two labels: ``normal" and ``attack", with a total of 449,921 pieces of data, of which 54,621 pieces are attack data.

\textbf{UNSW-NB15} is a public intrusion detection dataset~\cite{Moustafa2016}, which contains a total of 240,044 pieces of data, including 300,000 pieces of attack data, with 9 type attacks, i.e., analysis, backdoors, DoS, exploits, fuzzers, generic, reconnaissance and shellcode.

In SWaT dataset and UNSW-NB15 dataset, the data is is equipped with characteristics of large scale, irregularity, multiple features, temporal correlation and high dimensionality. Moreover, in the simulation experiment, we divide the selected SWaT data into a training set and a test set according to 8:2. This is because the sample distribution of the SWaT data set is uneven, and we need to include the corresponding attack data in the test set, so the selected test set has a large amount of data. Meanwhile, due to the large amount of UNSW-NB15 data, to facilitate the experiment, we randomly selected 25\% of the UNSW-NB15 dataset as experimental data, with a ratio of 10:1 for the training set and test set.

\subsection{Experimental criteria}
\label{sec:criteria}

In the simulation experiment, we use $Accuracy$, $Precision$, $Recall$, $F1\_score$ and false rejection rate ($FRR$) as experimental evaluation criteria to confirm the effectiveness of our proposed ARN-based intrusion detection method.

\subsection{Experimental results and analysis}

In this part, we develop a prototype implementation to verify the effectiveness of our proposed method. Specifically, we compared our ARN-based intrusion detection method with the GRU-based intrusion detection method~\cite{Xu2018}, AE-LSTM-based intrusion detection method~\cite{Mushtaq2022}, DR-AD-based intrusion detection method~\cite{KSood2023}, BiDLSTM-based intrusion detection method~\cite{ImranaY2021} and ICS-CAD-based intrusion detection method~\cite{Jadidi2023Multi}.

\subsubsection{Effectiveness evaluation of SWaT}

\label{subsubsec:SWaT}

In this simulation experiment, we verify the effectiveness of these methods based on industrial security dataset (SWaT) by using the experimental criteria described in Section~\ref{sec:criteria}.

\begin{figure}[t]
  \centering
  \includegraphics[width=3.5in]{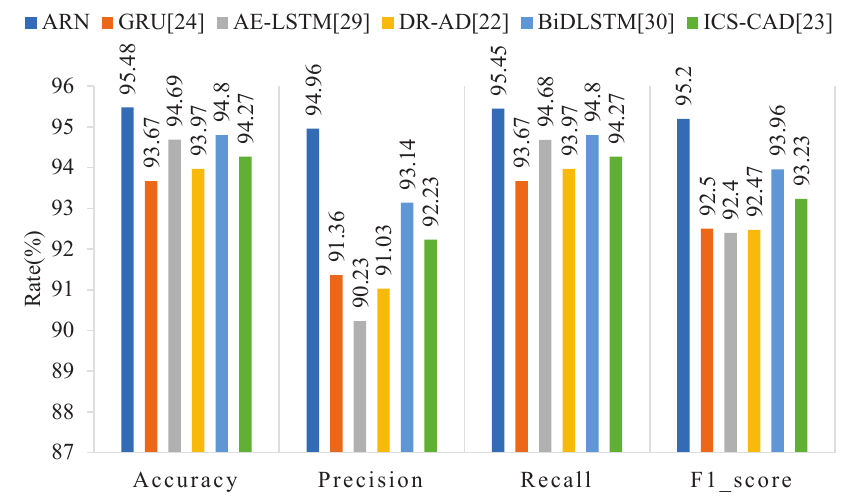}
  \caption{Comparative experiments under $Accuracy$, $Precision$, $Recall$ and $F1\_score$ of SWaT}
  \label{fig6}
\end{figure}

Firstly, we use $Accuracy$, $Precision $, $Recall $ and $F1\_score $ to conduct simulation experiments to further confirm the effectiveness of our method. Then, the experimental comparison results of these four indicators are depicted in Figure~\ref{fig6}. The SWaT dataset samples are extremely imbalanced, which makes the attack traffic difficult to be detected, resulting in similar indicator score differences. But our proposed method outperforms the other four existing methods in terms of $Accuracy$, $Precision$, $Recall$ and $F1\_score$, as shown in Figure~\ref{fig6}. That is, our proposed ARN method can provide more effective intrusion detection, thereby effectively improving the security of equipment in ICN.

\begin{figure}[t]
  \centering
  \includegraphics[width=3.5in]{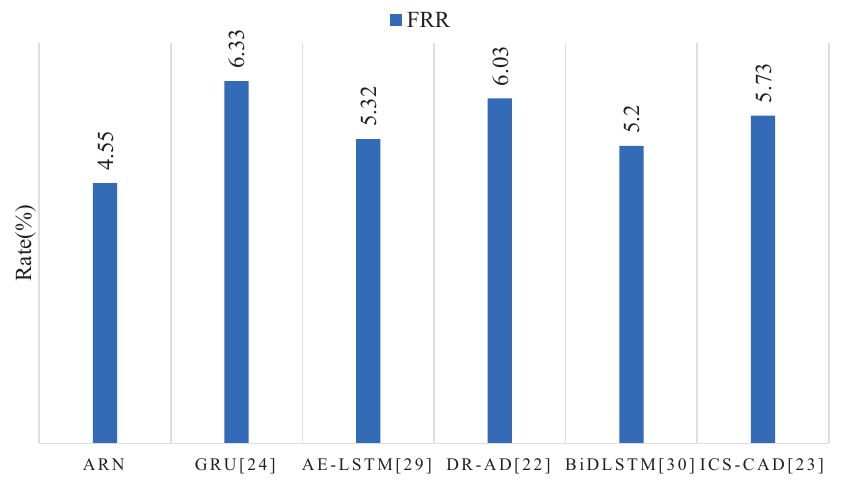}
  \caption{Comparative experiments under $FRR$ of SWaT}
  \label{fig7}
\end{figure}

Meanwhile, we also utilized the $FRR$ index to further confirm the effectiveness of our proposed method. Then, the experimental results are depicted in Figure~\ref{fig7}. We can know that our proposed method has obvious advantages in $FRR$. Therefore, we believe that our proposed method can improve the protection performance.

Overall, the SWaT dataset samples are extremely uneven, and it is difficult to successfully detect attack traffic. However, based on the evaluation indicators described in section~\ref{sec:criteria}, our proposed the ARN-based intrusion detection method has the best effect in all indicators. This is because for the GRU-based intrusion detection method ~\cite{Xu2018}, GRU has a gated unit conflict, which makes it difficult to effectively characterize the information content of the past moment, so the detection indicators are poor. For the AE-LSTM-based intrusion detection method ~\cite{Mushtaq2022}, it uses AE for feature extraction and then uses LSTM for detection, the detection accuracy is limited by the effect of feature extraction and the gated unit conflict of LSTM itself, and the loss of feature retention by the gated unit. For the DR-AD-based intrusion detection method~\cite{KSood2023}, its model structure itself cannot use the temporal nature of network traffic to characterize and learn network traffic, then the detection index is relatively poor. The BiDLSTM-based intrusion detection method~\cite{ImranaY2021} can use the temporal nature of network traffic to learn network data in both directions, but the detection results are still limited by the problems of LSTM itself. The ICS-CAD-based intrusion detection method~\cite{Jadidi2023Multi} is based on LSTM, so its detection accuracy is also limited by the inherent problems of LSTM. However, the ARN method effectively solves the conflict problem between gating units, so it can better manage the retention relationship between hidden state data at the past moment and data at the current moment. Therefore, compared with other existing methods, our proposed method can achieve the best intrusion detection results.

\subsubsection{Effectiveness evaluation of UNSW-NB15}

In this part, we conducted corresponding experiments on a general multi-class network intrusion detection data set, i.e., UNSW-NB15 to verify the versatility of our proposed method.

\begin{figure}[t]
  \centering
  \includegraphics[width=3.5in]{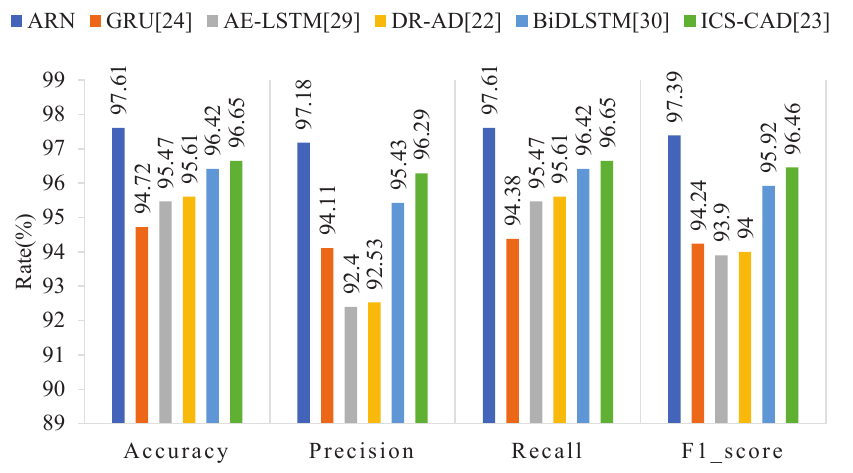}
  \caption{Comparative experiments under $Accuracy$, $Precision$, $Recall$ and $F1\_score$ of UNSW-NB15}
  \label{fig8}
\end{figure}

Figure~\ref{fig8} verifies our proposed method from $Accuracy$, $Precision$, $Recall$ and $F1\_score$. It can be seen that our proposed method has very obvious advantages on the UNSW-NB15 dataset.

\begin{figure}[t]
  \centering
  \includegraphics[width=3.5in]{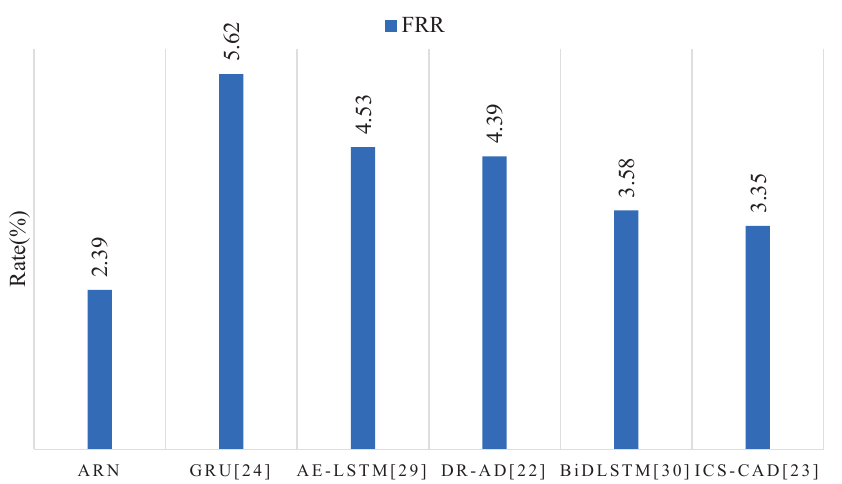}
  \caption{Comparative experiments under $FRR$ of UNSW-NB15}
  \label{fig9}
\end{figure}

In addition, the $FRR$ index in Figure~\ref{fig9} also further verifies the universality and effectiveness of our proposed method.

It can be seen that in the UNSW-NB15 dataset, the ARN-based intrusion detection method still has the best results in various indicators. The specific reasons are basically the same as the Section~\ref{subsubsec:SWaT} analysis. The existing RNN-based intrusion detection methods (AE-LSTM~\cite{Mushtaq2022}, GRU~\cite{Xu2018}, BiDLSTM~\cite{ImranaY2021} and ICS-CAD~\cite{Jadidi2023Multi}) have conflicts between gated units and the gated units have certain feature losses in the process of learning past moment information. Therefore, there are certain deficiencies in characterizing and learning past moment information, which leads to a decrease in detection accuracy. The DR-AD-based intrusion detection method~\cite{KSood2023} cannot use the temporal nature of network data to characterize and learn network traffic data, so the intrusion detection effect is poor. Meanwhile, both AE-LSTM and DR-AD have the operation of dimensionality reduction and detection, which will have the situation where the feature loss in the feature extraction process is superimposed on the detection process. It may make the original feature extraction unable to play the due effect and make the detection result worse. Because the ARN model does not have the problem of gated unit conflict, it can better characterize and learn past moment information, thereby effectively improving the intrusion detection results.

In summary, it can be seen that our proposed ARN-based intrusion detection method has good detection performance in both industrial security traffic and general network traffic.

\subsubsection{Efficiency Analysis}
\label{sec:efficiency}

\begin{figure}[t]
  \centering
  \includegraphics[width=3.5in]{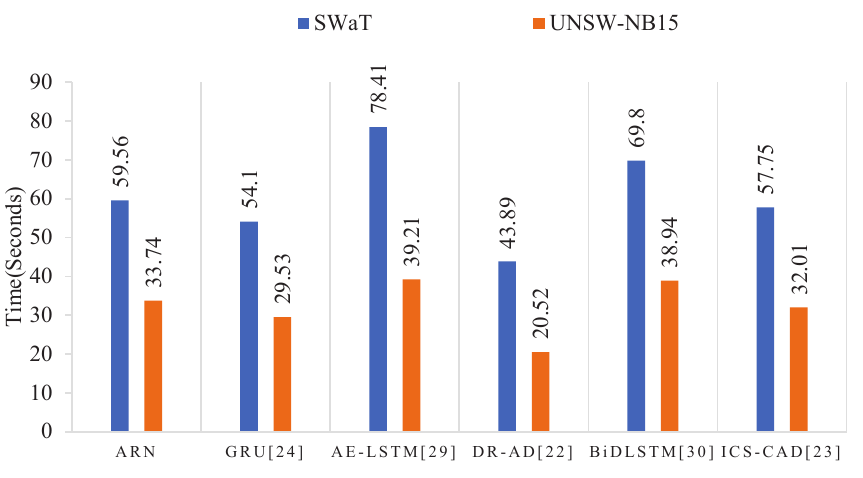}
  \caption{Test efficiency comparison}
  \label{fig10}
\end{figure}

In the following section, we analyze the performance of ARN-based intrusion detection method, GRU-based intrusion detection method~\cite{Xu2018}, AE-LSTM-based intrusion detection method~\cite{Mushtaq2022}, DR-AD-based intrusion detection method~\cite{KSood2023}, BiDLSTM-based intrusion detection method~\cite{ImranaY2021} and ICS-CAD-based intrusion detection method~\cite{Jadidi2023Multi} based on SWaT and UNSW-NB15 datasets. The comparison results are displayed in Figure~\ref{fig10}.

From Figure~\ref{fig10} it was found that the computational efficiency of our proposed ARN method is highly competitive and close to the GRU-based intrusion detection method~\cite{Xu2018}. The reason is that compared with GRU, ARN costs slightly more total computational overhead, and a detailed analysis can be found in Section~\ref{sec:analysis}. Meanwhile, the DR-AD-based intrusion detection method~\cite{KSood2023} has the lowest time overhead due to the simple BP network structure. The core of the ICS-CAD-based intrusion detection method~\cite{Jadidi2023Multi} is LSTM, which has an additional gate structure compared to GRU and is relatively complex. Therefore, its time cost is not much different from that of the method we proposed. In addition, the AE-LSTM-based intrusion detection method~\cite{Mushtaq2022} takes more computational time overhead because it composed of a two-part network structure and has more activation functions for nonlinear transformation. Moreover, the BiDLSTM-based intrusion detection method~\cite{ImranaY2021} needs to use two different LSTMs to consider the forward and reverse time series at the same time, resulting in more calculations.

In summary, the computational efficiency of our proposed ARN method is close to that of the GRU-based intrusion detection method~\cite{Xu2018}. Moreover, our proposed intrusion detection method has better accuracy. Therefore, the small difference in computational time overhead is absolutely acceptable and it can be adapted to efficient intrusion detection systems where existing methods have been applied.

\section{Conclusion and future work}
\label{sec:con}

In ICN, network traffic data is equipped with characteristics of large scale, irregularity, multiple features, temporal correlation and high dimensionality, resulting in low accuracy for the existing intrusion detection methods. In this paper, we studied the designing of efficient and accurate intrusion detection method for ICN. We firstly designed a novel RNN model, namely, ARN. Subsequently, we adopted ARN to build a novel intrusion detection method for ICN. By taking the advantage of ARN, our proposed method greatly improved the accuracy of intrusion detection. Finally, we implemented our proposed ARN-based intrusion detection method and provided a performance evaluation. Compared with some existing intrusion detection methods, the experimental results showed that the accuracies of our proposed method on the SWaT dataset and the UNSW-NB15 dataset reach 95.48\% and 97.61\%, respectively.

In practical application, ARN may have the same problem as other RNN methods, that is, in the process of characterizing and learning past information, ARN and existing RNN methods will eliminate some past moment information that is irrelevant to the current moment by calculating the association between past moment information and current input. However, this irrelevant information may be of great significance to some distant moment input in the future. This leads to a certain loss in ARN and existing RNN methods when characterizing and learning of temporal data. In addition, the network security managers not only want to detect the intrusion in real time but also want to know the network security situation at the next moment, thus taking some related defensive measures in advance. In this paper, we only studied the intrusion detection for ICN, which not involved in network security situation forecast. Hence, we will study the above limitations in the ARN model and research the security situation forecast for ICN in the future.


\begin{thebibliography}{00}


\bibitem{Galloway2012Introduction} B. Galloway and G. P. Hancke, ``Introduction to industrial control networks,'' \emph{IEEE Communications surveys and tutorials}, vol. 15, no. 2, pp. 860-880, 2012.
    
\bibitem{Cao2020Reliable} T, Cao, C. Xu, J. Du, Y. Li, H. Xiao and C. Gong, ``Reliable and efficient multimedia service optimization for edge computing-based 5G networks: Game theoretic approaches,'' \emph{IEEE Transactions on Network and Service Management}, vol. 17, no. 3, pp. 1610-1625, 2020.

\bibitem{Koursioumpas2021AIdriven} N. Koursioumpas, S. Barmpounakis, I. Stavrakakis and N. Alonistioti, ``AI-driven, context-aware profiling for 5G and beyond networks,'' \emph{IEEE Transactions on Network and Service Management}, vol. 19, no. 2, pp. 1036-1048, 2021.
    

\bibitem{Ding2019Asurvey} D. Ding, Q. L, Han, Z. Wang Z and X. Ge, ``A survey on model-based distributed control and filtering for industrial cyber-physical systems,'' \emph{IEEE Transactions on Industrial Informatics}, vol. 15, no. 5, pp. 2483-2499, 2019.


\bibitem{Lv2020Trust} Z. Lv \emph{et al., ``}Trustworthiness in industrial IoT systems based on artificial intelligence,'' \emph{IEEE Transactions on Industrial Informatics}, vol. 17, no. 2, pp. 1496-1504, 2020.


\bibitem{Du2023Afewshot} L. Du, Z. Gu, Y. Wang, L. Wang and Y. Jia, ``A Few-Shot Class-Incremental Learning Method for Network Intrusion Detection,'' \emph{IEEE Transactions on Network and Service Management}, vol. 21, no. 2, pp. 2389-2401, 2024.

\bibitem{He2024Reinforcement} M. He, X. Wang, P. Wei, L. Yang, Y. Teng and R. Lyu, ``Reinforcement learning meets network intrusion detection: a transferable and adaptable framework for anomaly behavior identification,'' \emph{IEEE Transactions on Network and Service Management}, vol. 21, no. 2, pp. 2477-2492, 2024.



\bibitem{Wu2021Tensor} Q. Wu, Z. Jiang, K. Hong, H. Liu, L. T. Yang and J. Ding et al, ``Tensor-based recurrent neural network and multi-modal prediction with its applications in traffic network management,'' \emph{IEEE Transactions on Network and Service Management}, vol. 18, no. 1, pp. 780-792, 2021.
    
\bibitem{Jha2020Recurrent} S. Jha, D. Prashar, H. V. Long and T. David, ``Recurrent neural network for detecting malware,'' \emph{computers and security}, vol. 99, pp. 102037, 2020.



\bibitem{Abdel2020DEEPIFS} M. Abdel-Basset \emph{et al., ``}Deep-IFS: Intrusion detection approach for industrial internet of things traffic in fog environment,'' \emph{IEEE Transactions on Industrial Informatics}, vol. 17, no. 11, pp. 7704-7715, 2020.


\bibitem{ChoK2014} K. Cho \emph{et al., ``}Learning phrase representations using RNN encoder-decoder for statistical machine translation,'' presented at the \emph{20\textsuperscript{th} Conference on Empirical Methods in Natural Language Processing (EMNLP 2014)}, Doha, Qatar, pp. 1724-1734, October 25-29, 2014.






\bibitem{Aitchison1982} J. Aitchison, ``The statistical analysis of compositional data,'' \emph{Journal of the Royal Statistical Society: Series B (Methodological)}, vol. 44, no. 2, pp. 139-160, 1982.


\bibitem{Zou2019} Y. Zou \emph{et al., ``}Complex network approaches to nonlinear time series analysis,'' \emph{Physics Reports}, vol. 787, pp. 1-97, 2019.

\bibitem{Dong2021As} S. Dong, P. Wang and K. Abbas, ``A survey on deep learning and its applications,'' \emph{Computer Science Review}, vol. 40, pp. 100379, 2021.




\bibitem{Wu2018} W. Wu \emph{et al., ``}Sliding window optimized information entropy analysis method for intrusion detection on in-vehicle networks,'' \emph{IEEE Access}, vol. 6, pp. 45233-45245, 2018.


\bibitem{Hu2020Detecting} Y. Hu \emph{et al., ``}Detecting stealthy attacks on industrial control systems using a permutation entropy-based method,'' \emph{Future Generation Computer Systems}, vol. 108, pp. 1230-1240, 2020.


\bibitem{Zhou2021Permutation} M. Zhou, Z. Zhang and L. Xie, ``Permutation entropy based detection scheme of replay attacks in industrial cyber-physical systems,'' \emph{Journal of the Franklin Institute}, vol. 358, no. 7, pp. 4058-4076, 2021.



\bibitem{Liao2022} D. Liao \emph{et al., ``}GE-IDS: an intrusion detection system based on grayscale and entropy,'' \emph{Peer-to-Peer Networking and Applications}, vol. 15, no. 3, pp. 1-14, 2022.



\bibitem{Liang2019An} W. Liang \emph{et al., ``}An industrial network intrusion detection algorithm based on multifeature data clustering optimization model,'' \emph{IEEE Transactions on Industrial Informatics}, vol. 16, no. 3, pp. 2063-2071, 2019.


\bibitem{Mittal2022A} H. Mittal \emph{et al., ``}A new intrusion detection method for cyber¨Cphysical system in emerging industrial IoT,'' \emph{Computer Communications}, vol. 190, pp. 24-35, 2022.




\bibitem{Bozdal2021} M. Bozdal, M. Samie and I. K. Jennions, ``WINDS: a wavelet-based intrusion detection system for controller area network (CAN),'' \emph{IEEE Access}, vol. 9, pp. 58621-58633, 2021.

\bibitem{Miao2020} K. Miao, X. Shi and W. A. Zhang, ``Attack signal estimation for intrusion detection in industrial control system,'' \emph{Computers and Security}, vol. 96, no. 101926, 2020.

\bibitem{Fouladi2016} R. F. Fouladi, C. E. Kayatas and E. Anarim, ``Frequency based DDoS attack detection approach using naive bayes classification,'' presented at the \emph{39\textsuperscript{th} International Conference on Telecommunications and Signal Processing (TSP)}, Austria, Vienna, pp. 104-107, June 27-29, 2016.




\bibitem{Ye2013} X. Ye, J. Lan and W. Huang, ``Network traffic anomaly detection based on self-similarity using FRFT,'' presented at the \emph{4\textsuperscript{th} International Conference on Software Engineering and Service Science}, Beijing, China, pp. 837-840, May 23-25, 2013.

\bibitem{Huang2020False} D. Huang, X. Shi and W. A. Zhang, ``False data injection attack detection for industrial control systems based on both time-and frequency-domain analysis of sensor data,'' \emph{IEEE Internet of Things Journal}, vol. 8, no. 1, pp. 585-595, 2020.









\bibitem{Li2020Robust} Y. Li et al. \emph{et al., ``}Robust detection for network intrusion of industrial IoT based on multi-CNN fusion,'' \emph{Measurement}, vol. 154, pp. 107450, 2020.

\bibitem{Abdelaty2021DAICS} M. Abdelaty, R. Doriguzzi-Corin and D. Siracusa, ``DAICS: A deep learning solution for anomaly detection in industrial control systems,'' \emph{IEEE Transactions on Emerging Topics in Computing}, vol. 10, no. 2, pp. 1117-1129, 2021.



\bibitem{Yang2019} A. Yang \emph{et al., ``}Design of intrusion detection system for internet of things based on improved BP neural network,'' \emph{IEEE Access}, vol. 7, pp. 106043-106052, 2019.



\bibitem{KSood2023} K. Sood \emph{et al., ``}Intrusion detection scheme with dimensionality reduction in next generation networks,'' \emph{IEEE Transactions on Information Forensics and Security}, vol. 15, pp. 965-979, 2023.


\bibitem{Jadidi2023Multi} Z. Jadidi, J. Hagemann and D. Quevedo, ``Multi-step attack detection in industrial control systems using causal analysis,'' \emph{Computers in Industry}, vol. 142, pp. 103741, 2022.



\bibitem{Xu2018} C. Xu \emph{et al., ``}An intrusion detection system using a deep neural network with gated recurrent units,'' \emph{IEEE Access}, vol. 6, pp. 48697-48707, 2018.


\bibitem{Zare2024A} F. Zare, P. Mahmoudi-Nasr and R. Yousefpour, ``A Real-Time Network Based Anomaly Detection in Industrial Control Systems,'' \emph{International Journal of Critical Infrastructure Protection}, pp. 100676, 2024. online



\bibitem{Vaswani2017} A. Vaswani \emph{et al., ``}Attention is all you need,'' presented at the \emph{31\textsuperscript{st} International Conference on Neural Information Processing Systems}, NY, United States, pp. 5998-6008, December 4-9, 2017.



\bibitem{JGoh2016} J. Goh \emph{et al., ``}A dataset to support research in the design of secure water treatment systems,'' presented at the \emph{Critical Information Infrastructures Security (CRITIS)} Paris, France, pp. 88-99, October 10¨C12, 2016.




\bibitem{Moustafa2016} N. Moustafa and J. Slay J, ``The evaluation of network anomaly detection systems: statistical analysis of the UNSW-NB15 data set and the comparison with the KDD99 data set,'' \emph{Information Security Journal: A Global Perspective}, vol. 25, no. 1-3, pp. 18-31, 2016.





\bibitem{Mushtaq2022} E. Mushtaq \emph{et al., ``}A two-stage intrusion detection system with auto-encoder and LSTMs,'' \emph{Applied Soft Computing}, vol. 121, pp. 108768, 2022.

\bibitem{ImranaY2021} Y. Imrana \emph{et al., ``}A bidirectional LSTM deep learning approach for intrusion detection,'' \emph{Expert Systems with Applications}, vol. 185, pp. 115524, 2021.






\end{thebibliography}
\end{document}